\documentclass[preprint,nobibnotes,prl,twocolumn,lengthcheck,10pt]{revtex4}%
\usepackage{amsfonts}
\usepackage{amsmath}
\usepackage{amssymb}
\usepackage{graphicx}%
\begin{document}
\preprint{UATP/04-06}
\title{Lack of Stability in the Stillinger-Weber Analysis, and a Stable Analysis of
the Potential Energy Landscape}
\author{P. D. Gujrati}
\affiliation{The Department of Physics, The Department of Polymer\ Science, The University
of\ Akron, Akron, Ohio 44325}
\date{\today }

\pacs{PACS number}

\begin{abstract}
We examine the Stillinger-Weber \ analysis of the potential energy landscape
for its stability and conclude that it does not provide a stable description
of the system as the free energy slope and curvature vanish simultaneously. An
alternative analysis developed recently by us involving complexity provides a
stable description with complexity a monotonic increasing function of temperature.

\end{abstract}
\maketitle

It is well known that most supercooled liquids (SCL) become viscous when their
configurational entropy becomes negligible as they are cooled, provided the
corresponding crystal is not allowed to nucleate.\ Our current understanding
of glassy behavior is still far from complete, even after many decades of
continuous investigation. In order to better understand the flow properties of
viscous fluids, Goldstein proposed the potential energy landscape (PEL)
picture using \emph{classical\ canonical ensemble }\cite{Goldstein,Goldstein1}
to qualitatively discuss an interesting but sufficiently tractable scheme to
study SCL and the glassy states by drawing attention to the potential energy
minima (having the energy $\mathcal{E)},$ to be called basin minima (BM) in
the following. The landscape picture with its BM's plays a pivotal role not
only in the thermodynamics of viscous fluids at \emph{low temperatures}\ but
in many disparate fields like glasses, proteins and clusters \cite{Wales}, and
has established itself as an important thermodynamic approach in theoretical
physics. Thus, it is highly desirable to understand the significance of this
approach. Stillinger and Weber (SW) extended the work of Goldstein to higher
temperatures by carrying out a formal analysis in terms of the minima energies
$\mathcal{E}$ \cite{Stillinger,Stillinger1}. Their analysis has given rise to
a considerable amount of literature\ in recent years; for a partial list, see
\cite{Stillinger,Stillinger1,Stillinger2,ISliterature}. Many of the numerous
numerical evidence \cite{ISliterature} appear to be consistent with
Goldstein's seminal ideas \cite{Goldstein}.

In this work, we study the stability of the SW analysis, which seems not to
have been investigated in the literature. There are two different conditions
for the stability of a thermodynamic theory \cite{Landau}. The first one is
the vanishing of the slope of the free energy function and is commonly
discussed in the literature. The other condition is of a strictly positive
curvature of the free energy function at the point where the first condition
is met. This does not appear to have been ever discussed. To our surprise, we
find that the free energy function in the SW analysis has a zero curvature.
Thus, the SW analysis does not give rise to a stable description of the system
and must be replaced by other self-consistent approaches. We have recently
developed such an approach \cite{GujratiSemerianov,Gujrati}, which borrows the
concept of complexity developed for spin glasses \cite{Parisi, Parisi1}. The
new analysis has no problem with stability and is consistent.

\textbf{Conventional Approach. }The canonical PF $Z(T)$ for a system
of$\ N\ $particles in a volume $V$ is%

\begin{equation}
Z(T)\equiv\sum_{E}W(E)e^{-\beta E}, \label{StandardPF}%
\end{equation}
where $W(E)$ represents the number of configurations of energy $E$ and defines
the microcanonical entropy $S(E)\equiv\ln$ $W(E),$ and $\beta$ is the inverse
temperature in the units of the Boltzmann constant$.$ The value of $Z(T)$ for
a macroscopic system, which is what we consider here, is determined by the
dominant term in the sum, which is located at the equilibrium energy
$\overline{E}\equiv E(T)$: $Z(T)\cong W[\overline{E}]\exp(-\beta\overline
{E}).$\ The determination of $\overline{E}$\ for a macroscopic system is
simplified by noting that $E$ is almost a continuous variable for a
macroscopic system. In terms of $S(E),$ $\overline{E}$ is given by the
\emph{location of the minimum} of the free energy function $F(T,E)=E-TS(E)$ at
fixed $T.$ In equilibrium, the entropy $S(T)\equiv S(E=\overline{E})$ and free
energy $F(T)\equiv F(T,\overline{E})=E(T)-TS(T)$\ become functions only of
$T.$ The conditions for the minimum are $[\partial F(T,E)/\partial E]_{T}=0,$
and $[\partial^{2}F(T,E)/\partial E^{2}]_{T}>0$ leading to
\begin{equation}
\left[  \partial S(E)/\partial E\right]  _{E=E(T)}=\beta,\text{ }\partial
S(T)/\partial T>0, \label{EquilibriumE}%
\end{equation}
which are always satisfied because of a non-negative heat capacity. With the
use of (\ref{EquilibriumE}), we immediately conclude
\begin{equation}
T(\partial S(T)/\partial T)=\partial E(T)/\partial T, \label{FirstLaw}%
\end{equation}
which is consistent with the first law of thermodynamics at constant $V,$ and
$N.$

At a given temperature $T,$ only those configurations that have the energy
$E=\overline{E}$ (or within a narrow width around it, depending on the heat
capacity; we will neglect this width here) determine the thermodynamics
through the entropy $S(T)$. All energies other than $\overline{E}$ and,
therefore, all configurations not included in $W(\overline{E})$ are irrelevant
at $T.$ Thus, thermodynamics is highly \emph{selective}. This will remain true
even in the landscape picture, where the equilibrium states will have the same
energy regardless of which basin they belong to. Thus,\emph{ }$\overline
{E}=E(T)$\emph{ cannot depend explicitly on the basin minima energy
}$\mathcal{E}$:$(\partial\overline{E}/\partial\mathcal{E})_{T}=0.$

\textbf{Goldstein's Approximate\ Analysis. }In his analysis, Goldstein has
listed two conjectures that were common in the field \cite{Goldstein1} at the
time: the basin PF $z_{\text{b}}(T)$ is (i) independent of the basin's minimum
energy $\mathcal{E},$ and (ii) insensitive to the basins being explored.
Utilizing these assumptions, Goldstein has expressed the PF as a product
\cite{Goldstein1}\ of the basin and BM PF$^{\prime}$%
\begin{equation}
Z(T)=z_{\text{b}}(T)Z_{\text{BM}}(T); \label{GoldsteinPart}%
\end{equation}
here $z_{\text{b}}$ for a given basin is defined by considering shifted
energies $E-\mathcal{E}$ with respect to the minimum energy $\mathcal{E}$ of
that basin; see also Schulz \cite{Schulz}. Goldstein has emphasized that basin
anharmonicity or the curvature at its minimum \cite{Angell} may be very
important. These are included in $z_{\text{b}},$ so that it is determined by
the entire basin topology$.$ According to Goldstein, all equilibrium basins
have the same equilibrium basin free energy $f_{\text{b}}(T)\equiv-T\ln$
$z_{\text{b}}$. The BM-PF is defined \cite{Goldstein1,Schulz} as%
\begin{equation}
Z_{\text{BM}}(T)=\mathbf{\sum_{\mathcal{E}}}N_{\text{BM}}(\mathcal{E}%
)e^{-\beta\mathcal{E}}. \label{GoldsteinConfPF}%
\end{equation}
Here, $N_{\text{BM}}(\mathcal{E})$ represents the number of basins whose BM
are at energy $\mathcal{E}.$ The equilibrium BM energy $\overline{\mathcal{E}%
}=\mathcal{E}(T)$\ is the value of $\mathcal{E}$\ at which the summand in
(\ref{GoldsteinConfPF}) is maximum. The conditions for the maximum in terms of
the BM entropy $S_{\text{BM}}(\mathcal{E})\equiv\ln N_{\text{BM}}%
(\mathcal{E})$ are given by
\begin{equation}
\lbrack\partial S_{\text{BM}}(\mathcal{E})/\partial\mathcal{E}]_{\mathcal{E}%
=\overline{\mathcal{E}}\ }=\beta,\text{ }\partial\overline{\mathcal{E}%
}/\partial T>0, \label{Goldstein1}%
\end{equation}
which are the standard conditions of equilibrium; compare with
(\ref{EquilibriumE}). Thus, the analysis is completely stable in this
approximation. It is clear that the BM description proposed by Goldstein
ensures that $\overline{\mathcal{E}}$\emph{ is a monotonic increasing function
of }$T.$ Since this approximation is expected to be good at low temperatures,
we expect $\overline{\mathcal{E}}$ to be monotonic increasing there. But it
need not be true at all temperatures as shown below, thereby limiting the
usefulness of the BM-description at all temperatures that has been formally
adopted by Stillinger and Weber to which we now turn.

\textbf{SW\ Analysis. }A basin\ is indexed by $j$, and the lowest and highest
basin energies are denoted by $\mathcal{E}_{j}$, and $\mathcal{E}%
_{j}^{^{\prime}},$ respectively, so that the basin does not exist outside the
energy range $\Delta_{j}E\equiv$ $(\mathcal{E}_{j},\mathcal{E}_{j}^{^{\prime}%
})$. Let $W_{j}(E)$ ($E\in\Delta_{j}E$) represent the number of distinct
configurations of energy $E$ in the $j$-th basin and introduce the entropy
$S_{j}(E)\equiv\ln W_{j}(E)$. We now introduce the \emph{shifted} PF
\begin{equation}
z_{j}(T)\equiv\underset{E\in\Delta_{j}E}{\sum}W_{j}(E)e^{-\beta(E-\mathcal{E}%
_{j})}\; \label{ShftPF}%
\end{equation}
of the $j$-th basin and the free energy function $f_{j}(\mathcal{E}%
_{j},E,T)\equiv E-\mathcal{E}_{j}-TS_{j}(E),$ determined by the general
summand in (\ref{ShftPF}). The form of $f_{j}(\mathcal{E}_{j},E,T)$ assumes
that $\mathcal{E}_{j},E$ are \emph{independent}, which is consistent with what
was said above about the energy $\overline{E}$ of equilibrium
configurations\ and its independence from the basins to which they belong. The
conditions for the minimum of $f_{j}(\mathcal{E}_{j},E,T)$\ at $E=\overline
{E}_{j}\equiv E_{j}(T)$ are
\begin{subequations}
\begin{align}
(\partial f_{j}(\mathcal{E}_{j},E,T)/\partial E\left)  _{\mathcal{E}_{j}%
,T}\right\vert _{E=\overline{E}_{j}}  &  =0,\text{ }\label{BasinEquilibrium0}%
\\
(\partial^{2}f_{j}(\mathcal{E}_{j},E,T)/\partial E^{2}\left)  _{\mathcal{E}%
_{j},T}\right\vert _{E=\overline{E}_{j}}  &  >0, \label{BasinEquilibrium00}%
\end{align}
and simplify to
\end{subequations}
\begin{equation}
(\partial S_{j}(E)/\partial E\left)  _{\mathcal{E}_{j}}\right\vert
_{E=\overline{E}_{j}}=\beta,\text{ }\partial E_{j}(T)/\partial T>0.
\label{BasinEquilibrium}%
\end{equation}
Both conditions are always met. The \emph{average} basin energy $E_{j}(T)$
determines the average entropy $S_{j}(T)\equiv S_{j}(E=\overline{E}_{j}),$ so
that $f_{j}(T)\equiv-T\ln z_{j}(T)=E_{j}(T)-\mathcal{E}_{j}-TS_{j}(T)$
represents the basin free energy. We wish to emphasize that $E_{j}%
(T),S_{j}(T),$ and $f_{j}(T)$ do \emph{not} represent equilibrium quantities
yet; the latter are determined only after $Z(T)$ is evaluated. (If each basin
is treated as representing an independent system in a formal sense, then these
quantities do represent equilibrium values for the particular basin.)

We now group basins, indexed by $j(\lambda)$, into basin classes (BC)
$\mathcal{B}_{\lambda}$, indexed by $\lambda$, so that all basins in a class
have the same BM energy $\mathcal{E}=\mathcal{E}_{\lambda}.$ The basins in a
class do \emph{not} have to be close in the configuration space. The number of
basins in $\mathcal{B}_{\lambda}$ is $N_{\text{BM}}(\mathcal{E}_{\lambda}),$
and the corresponding BM entropy is $S_{\text{BM}}(\mathcal{E}_{\lambda
})\equiv\ln N_{\text{BM}}(\mathcal{E}_{\lambda})$. Let%
\begin{equation}
Z_{\lambda}(T)\equiv\underset{j\in j(\lambda)}{\sum}z_{j}(T),\text{
\ }z_{\lambda}\equiv Z_{\lambda}(T)/N_{\text{BM}}(E_{\lambda}),
\label{BasinClassPF}%
\end{equation}
denote the shifted and the mean shifted basin $\mathcal{B}_{\lambda}$-PF,
respectively, so that
\begin{subequations}
\begin{align}
Z(T)  &  \equiv\sum_{\lambda}e^{-\beta\mathcal{E}_{\lambda}+{}S_{\text{BM}%
}(\mathcal{E}_{\lambda})}z_{\lambda}(T),\,\;\label{ISTotPF1}\\
\mathcal{E}(T)  &  \equiv\sum_{\lambda}\mathcal{E}_{\lambda}e^{-\beta
\mathcal{E}_{\lambda}+{}S_{\text{BM}}(\mathcal{E}_{\lambda})}z_{\lambda
}(T)/Z(T). \label{ISTotPF2}%
\end{align}
Here, $\overline{\mathcal{E}}=\mathcal{E}(T)$\ represents the equilibrium BM
energy. It is easy to see that $\partial\overline{\mathcal{E}}/\partial T$\ is
a cross-correlation so that it need not have a unique sign \cite{Gujrati}. The
equilibrium free energy, entropy and energy are\ $F(T)=-T\ln Z(T),$
$S(T)=-\partial F/\partial T$ and $E(T)=F(T)+TS(T),$ respectively$.$

\textbf{SW} \textbf{Assumption. }In the SW analysis, the sum over $\lambda$ in
(\ref{ISTotPF1},\ref{ISTotPF2}) is replaced by a sum over the BM energy
$\mathcal{E}_{\lambda}$ by assuming that $z_{\lambda}$ depends explicitly on
$\mathcal{E}_{\lambda}$ in addition to $T:$ $z_{\lambda}=z_{\lambda
}(\mathcal{E}_{\lambda},T).$ This issue has been examined earlier by us
\cite{GujratiSemerianov,Gujrati}. Here, we pursue the consequence of this
\emph{assumption }for the stability of this approach\emph{.} Let $f_{\lambda
}(\mathcal{E}_{\lambda},T)=-T\ln z_{\lambda}(\mathcal{E}_{\lambda},T)$ be the
mean free energy resulting from the mean basin PF $z_{\lambda}$ and
$S_{\lambda}(\mathcal{E}_{\lambda},T)=-[\partial f_{\lambda}(\mathcal{E}%
_{\lambda},T)/\partial T\mathcal{]}_{\mathcal{E}_{\lambda}}$ the mean basin
entropy$.$\ From the form of $Z_{\lambda}(T)$ and $z_{\lambda}(\mathcal{E}%
_{\lambda},T),$\ it is obvious that we can rewrite $z_{\lambda}(\mathcal{E}%
_{\lambda},T)$\ as follows:%
\end{subequations}
\begin{equation}
z_{\lambda}(\mathcal{E}_{\lambda},T)=\underset{E\in\Delta_{\lambda}E}{\sum
}W_{\lambda}(E)e^{-\beta(E-\mathcal{E}_{\lambda})-S_{\text{BM}}(\mathcal{E}%
_{\lambda})}, \label{MeanBCPF}%
\end{equation}
where $W_{\lambda}(E)$ represents the number of configurations of energy
$E$\ that belong to $\mathcal{B}_{\lambda}\ $and indicated by $E\in
\Delta_{\lambda}E.$\ Introducing $S_{\lambda}(\mathcal{E}_{\lambda}%
,E)\equiv\ln W_{\lambda}(E)-S_{\text{BM}}(\mathcal{E}_{\lambda}),$ we find
that the general summand in (\ref{MeanBCPF}) determines a mean free energy
function $f_{\lambda}(\mathcal{E}_{\lambda},E,T)\equiv E-\mathcal{E}_{\lambda
}-TS_{\lambda}(\mathcal{E}_{\lambda},E)$ whose minimization with respect to
$E$ at fixed $\mathcal{E}_{\lambda},$ and $T$ determines the mean free energy
$f_{\lambda}(\mathcal{E}_{\lambda},T).$\ Let $\overline{E}_{\lambda
}=E_{\lambda}(T)$\ denote the location of the minimum, the conditions for
which are exactly the same as for $f_{j}(\mathcal{E}_{j},E,T)$ above, except
that the index $j$ is replaced by $\lambda$ and $S_{\lambda}(\mathcal{E}%
_{\lambda},E)$ is a two-variable function. Again, $\mathcal{E}_{\lambda},$ and
$E$ are treated as two independent variables for the minimization to be
carried out. The conditions for minimization are
\begin{equation}
(\partial S_{\lambda}(\mathcal{E}_{\lambda},E)/\partial E\left)
_{\mathcal{E}_{\lambda}}\right\vert _{E=\overline{E}_{\lambda}}=\beta,\text{
}\partial E_{\lambda}(T)/\partial T>0. \label{meanBCCond}%
\end{equation}
The mean free energy and entropy are given by $f(\mathcal{E}_{\lambda
},T)\equiv$ $f_{\lambda}(\mathcal{E}_{\lambda},\overline{E}_{\lambda}%
,T)\equiv\overline{E}_{\lambda}-\mathcal{E}_{\lambda}-TS_{\lambda}%
(\mathcal{E}_{\lambda},\overline{E}_{\lambda})$ and $S_{\lambda}%
(\mathcal{E}_{\lambda},T)=S_{\lambda}(\mathcal{E}_{\lambda},\overline
{E}_{\lambda}),$ respectively. It is easy to see that
\begin{equation}
S_{\lambda}(\mathcal{E}_{\lambda},\overline{E}_{\lambda})=-[\partial
f_{\lambda}(\mathcal{E}_{\lambda},T)/\partial T\mathcal{]}_{\mathcal{E}%
_{\lambda}} \label{BasinFreeEnergyRealtion}%
\end{equation}
as expected, where we must use%
\begin{equation}
(\partial\overline{E}_{\lambda}/\partial T)=T[\partial S_{\lambda}%
(\mathcal{E}_{\lambda},\overline{E}_{\lambda})/\partial T]_{\mathcal{E}%
_{\lambda}}, \label{FirstLawBasinEntropy}%
\end{equation}
which follows immediately from the first condition in (\ref{meanBCCond}).
Since $\overline{E}_{\lambda}$ is independent of $\mathcal{E}_{\lambda}$ at
fixed $T,$ we can differentiate $f_{\lambda}(\mathcal{E}_{\lambda}%
,\overline{E}_{\lambda},T)$ with respect to $\mathcal{E}_{\lambda}$ to obtain%
\begin{equation}
(\partial S(\mathcal{E},T)/\partial\mathcal{E})_{T}=-\beta\lbrack1+(\partial
f(\mathcal{E},T)/\partial\mathcal{E)}_{T}], \label{meanSfrelation}%
\end{equation}
where fixed $T$ means keeping $\overline{E}_{\lambda}$ and $T$ fixed
simultaneously, and where we have suppresed $\lambda$ to treat $\mathcal{E}$ a variable.

\textbf{Zero-Slope Condition. }Because of the assumed $\mathcal{E}$-dependence
of $z_{\lambda}(T)$, the general summand in $(\ref{ISTotPF1},\ref{ISTotPF2})$
becomes an explicit function of $\mathcal{E},$\ and we can minimize the
corresponding free energy function $F_{\text{B}}(\mathcal{E},T)\equiv
\mathcal{E}+f(\mathcal{E},T)-TS_{\text{BM}}(\mathcal{E})$ with respect to
$\mathcal{E}$ at fixed\emph{\ }$T$ to determine $Z(T)$.\ The minimum of
$F_{\text{B}}(\mathcal{E},T)$ is given by the conditions $[\partial
F_{\text{B}}(\mathcal{E},T)/\partial\mathcal{E}]_{T}=0,$ and $[\partial
^{2}F_{\text{B}}(\mathcal{E},T)/\partial\mathcal{E}^{2}]_{T}>0.$ The first
condition is satisfied at the equilibrium BM-energy $\overline{\mathcal{E}%
}\mathcal{=\mathcal{E}(T)=E}(T),$ see (\ref{ISTotPF2}). It is also given by
the solution of
\begin{equation}
\partial S_{\text{BM}}(\mathcal{E})/\partial\mathcal{E}=\beta\lbrack
1+(\partial f(\mathcal{E},T)/\partial\mathcal{E)}_{T}], \label{ISTMaxCon}%
\end{equation}
and determines the equilibrium free energy $F(T)\equiv$ $F_{\text{B}%
}(\overline{\mathcal{E}},T)$, BM-entropy $S_{\text{BM}}(T)\equiv$
$S_{\text{BM}}(\mathcal{E}=\overline{\mathcal{E}}),$ mean basin free energy
$f_{\text{b}}(T)=f(\overline{\mathcal{E}},T)$ and mean basin entropy
$S_{\text{b}}(T)=S(\overline{\mathcal{E}},T)=-[\partial f(\overline
{\mathcal{E}},T)/\partial T\mathcal{]}_{\mathcal{E}};$ see
(\ref{BasinFreeEnergyRealtion}). The equilibrium mean basin energy
$E_{\text{b}}(T)$ is obtained by the fundamental relation $E_{\text{b}%
}(T)-\overline{\mathcal{E}}=f_{\text{b}}(T)+TS_{\text{b}}(T).$ It is easy to
see that the form of the equilibrium free energy $F(T)=f(\overline
{\mathcal{E}},T)+\mathcal{E(}T\mathcal{)}-TS_{\text{BM}}(\overline
{\mathcal{E}})$\ is the same as the free energy obtained by Goldstein in
(\ref{GoldsteinPart}), except that the equations determining the equilibrium
BM-energy are different; compare (\ref{Goldstein1}) and (\ref{ISTMaxCon}). The
two conditions become identical if $f$ is taken to be independent of
$\mathcal{E},$ as was assumed by Goldstein. Comparing (\ref{ISTMaxCon}) with
(\ref{meanSfrelation}) applied at $\mathcal{E}=\overline{\mathcal{E}},$ we
obtain an interesting relation
\begin{equation}
\partial S_{\text{BM}}(\overline{\mathcal{E}})/\partial\overline{\mathcal{E}%
}+(\partial S(\overline{\mathcal{E}},T)/\partial\overline{\mathcal{E}})_{T}=0,
\label{BC-BMEntropyRelation}%
\end{equation}
which will play a very important role in the following when we investigate the
stability of this approach.

The entropy $S(T)$ can now be obtained by using the relation $S(T)=-\partial
F(T)\partial T.$ We immediately find that $S(T)=S(\overline{\mathcal{E}%
},T)+S_{\text{BM}}(\overline{\mathcal{E}}).$ The equilibrium energy given by
$F(T)+TS(T),$ thus, turns out to be $E_{\text{b}}(T)$ introduced above. From
the conventional analysis, this energy was identified as $E(T).$ Thus,
\[
E_{\text{b}}(T)\equiv E(T).
\]
Now, we apply (\ref{FirstLawBasinEntropy}) at $\overline{E}_{\lambda
}=E_{\text{b}}(T)\equiv E(T)$ to find%
\begin{equation}
(\partial E(T)/\partial T)=T[\partial S(\overline{\mathcal{E}},\overline
{E})/\partial T]_{\overline{\mathcal{E}}}. \label{FirstLaw2}%
\end{equation}
It should be noted that the entropy derivative on the right-hand side is the
intrabasin change in the basin entropy with $T$ without leaving the basin
(fixed $\overline{\mathcal{E}}$). Comparing this with (\ref{FirstLaw}), we
find that the right hand side in both equations must be the same. This can
only happen if (\ref{BC-BMEntropyRelation}) is fulfilled; we assume that
$(\partial\overline{\mathcal{E}}/\partial T)\neq0.$ This provides another
justification for the validity of (\ref{BC-BMEntropyRelation}), and is merely
a consequence of the first condition of stability.

\textbf{Curvature Condition. }We now proceed to discuss the second condition
for minimization. This condition of stability at $\mathcal{E}=\overline
{\mathcal{E}}$ reads
\begin{equation}
(\partial^{2}f(\overline{\mathcal{E}},T)/\partial\overline{\mathcal{E}}%
^{2}\mathcal{)}_{T}-T(\partial^{2}S_{\text{BM}}(\overline{\mathcal{E}%
})/\partial\overline{\mathcal{E}}^{2}\mathcal{)}>0. \label{ISTMaxCon1}%
\end{equation}
We differentiate (\ref{ISTMaxCon}) at arbitrary $\mathcal{E},$ which yields%
\begin{align*}
\frac{\partial^{2}S_{\text{BM}}(\mathcal{E})}{\partial\mathcal{E}^{2}}  &
=\frac{\partial\beta}{\partial\mathcal{E}}[1+(\frac{\partial f(\mathcal{E}%
,T)}{\partial\mathcal{E}}\mathcal{)}_{T}]\mathcal{+}\\
&  \beta\lbrack(\frac{\partial^{2}f(\mathcal{E},T)}{\partial\mathcal{E}^{2}%
}\mathcal{)}_{T}+\frac{\partial^{2}f(\mathcal{E},T)}{\partial T\partial
\mathcal{E}}\frac{\partial T}{\partial\mathcal{E}}].
\end{align*}
We now set $\mathcal{E}=\overline{\mathcal{E}}$ and use it in
(\ref{ISTMaxCon1})\ to finally obtain the condition to be
\begin{equation}
\mathcal{[}\partial S_{\text{BM}}(\overline{\mathcal{E}})/\partial
\overline{\mathcal{E}}-\partial^{2}f(\overline{\mathcal{E}},T)/\partial
\overline{\mathcal{E}}\partial T](\partial T/\partial\overline{\mathcal{E}%
})>0, \label{ISTMaxCon2}%
\end{equation}
where we have used (\ref{ISTMaxCon}). Applying (\ref{BasinFreeEnergyRealtion})
at equilibrium, we obtain $S(\overline{\mathcal{E}},T)=-[\partial
f(\overline{\mathcal{E}},T)/\partial T\mathcal{]}_{\overline{\mathcal{E}}}.$
Thus, the numerator in (\ref{ISTMaxCon2}) can be reduced to $\partial
S_{\text{BM}}(\overline{\mathcal{E}})/\partial\overline{\mathcal{E}}+[\partial
S(\overline{\mathcal{E}},T)/\partial\overline{\mathcal{E}}]_{T}.$ [The
numerator can also be alternatively expressed as $\partial S(T)/\partial
\overline{\mathcal{E}}-[\partial S(\overline{\mathcal{E}},T)/\partial
T]_{\overline{\mathcal{E}}}/(\partial\overline{\mathcal{E}}/\partial T)$ where
$S(T)=S(\overline{\mathcal{E}},T)+S_{\text{BM}}(\overline{\mathcal{E}})]$.
Thus, the second condition of stability reads%
\begin{equation}
\text{\ \ }[\partial S_{\text{BM}}(\overline{\mathcal{E}})/\partial
\overline{\mathcal{E}}+[\partial S(\overline{\mathcal{E}},T)/\partial
\overline{\mathcal{E}}]_{T}](\partial T/\partial\overline{\mathcal{E}})>0,
\label{ISTMaxCon4}%
\end{equation}
which can never be satisfied in view of (\ref{BC-BMEntropyRelation}) unless
$(\partial\overline{\mathcal{E}}/\partial T)=0.$ Since it is evident from
(\ref{ISTotPF2}) that $(\partial\overline{\mathcal{E}}/\partial T)\neq0$\ in
general, we conclude that the curvature of the free energy function at
$\mathcal{E}=\overline{\mathcal{E}}$ must vanish on account of the first
condition of stability. Thus, we have finally shown that the SW analysis is
internally inconsistent and fails to provide a stable description of the system.

\textbf{Complexity Approach. }We provide an alternative approach
\cite{Gujrati} which, as we show below, turns out to be a consistent and
stable approach. We consider the \emph{unshifted} basin partition function
$Z_{j}(T)\equiv e^{\beta\mathcal{E}_{\lambda}}z_{j}(T)$ and introduce the
unshifted basin free energy $\varphi_{j}(T)=-T\ln Z_{j}(T).$ The conditions of
stability for the basin free energy $\varphi_{j}(E,T)\equiv E-TS_{j}(E)$ are
given in (\ref{BasinEquilibrium}). The basin free energy $\varphi_{j}%
(T)$\ varies from basin to basin and represents a \emph{family} of functions,
one for each $j$. Let $\mathcal{N}(\varphi,T)$ represent the number of basins
having the same free energy $\varphi$ for a given $T$ and rewrite
(\ref{StandardPF}) as%
\begin{equation}
Z(T)\equiv\sum_{\varphi}\mathcal{N}(\varphi,T)e^{-\beta\varphi}.
\label{StandardPF1}%
\end{equation}
The \emph{complexity} is defined by $\mathcal{S}(\varphi,T)\equiv
\ln\mathcal{N}(\varphi,T),$ in terms of which the conditions of stability at
$\varphi=\overline{\varphi}_{\text{b}}=\varphi_{\text{b}}(T)$
\begin{equation}
\left(  \partial\mathcal{S}(\varphi,T)/\partial\varphi)_{T}\right\vert
_{\varphi=\overline{\varphi}_{\text{b}}\text{\ \ \ }}=\beta,[\partial
^{2}\mathcal{S}(\overline{\varphi}_{\text{b}},T)/\partial\overline{\varphi
}_{\text{b}}^{2}]_{T}<0. \label{Equilibriumphi}%
\end{equation}
The equilibrium complexity $\overline{\mathcal{S}}(T)$ is given by
$\mathcal{S}(\overline{\varphi}_{\text{b}},T)$ evaluated at $\varphi
=\overline{\varphi}_{\text{b}}.$ We consider the case so that $\mathcal{S}%
(\varphi,T)$ can be inverted at fixed $T$ to express $\varphi$ as a function
of $\mathcal{S},T:\varphi=\varphi(\mathcal{S},T).$ Thus,%
\begin{equation}
d\varphi(\mathcal{S},T)=\left(  \partial\varphi/\partial\mathcal{S}\right)
_{T}d\mathcal{S+}\left(  \partial\varphi/\partial T\right)  _{\mathcal{S}}dT.
\label{DifferentialPhi}%
\end{equation}
At equilibrium, $\overline{\varphi}_{\text{b}}=\varphi(\overline{\mathcal{S}%
},T),$ and the coefficient of the first term becomes $T$ according to the
first relation in (\ref{Equilibriumphi}). In general, the coefficient of the
second term is the negative basin entropy: $S(\varphi,T)=-(\partial
\varphi(\mathcal{S},T)/\partial T)_{\mathcal{S}};$ compare with
(\ref{BasinFreeEnergyRealtion})$.$ Let us introduce $\sigma(\varphi,T)=\left(
\partial\mathcal{S}(\varphi,T)/\partial\varphi\right)  _{T}$ so that
$\sigma(\overline{\varphi}_{\text{b}},T)=\beta$ at equilibrium$.$ From
(\ref{DifferentialPhi}), we find%
\begin{align}
\left(  \partial\mathcal{S}(\varphi,T)/\partial T\right)  _{\varphi}  &
=-\partial\mathcal{S}(\varphi,T)/\partial\varphi)_{T}\left(  \partial
\varphi(\mathcal{S},T)/\partial T\right)  _{\mathcal{S}}\nonumber\\
&  =\sigma(\varphi,T)S(\varphi,T). \label{ComplexityCoefiiciet1}%
\end{align}
We differentiate $\sigma(\varphi,T)$ with respect to $T$ at constant $\varphi
$\ and use the above equation to obtain%
\begin{align*}
(\partial\sigma(\varphi,T)/\partial T)_{\varphi}  &  =\partial^{2}%
\mathcal{S}(\varphi,T)/\partial\varphi\partial T\\
&  =(\partial\lbrack\sigma(\varphi,T)S(\varphi,T)]/\partial\varphi)_{T},
\end{align*}
which is used to calculate
\begin{align*}
\partial\sigma(\varphi,T)/\partial T  &  =(\partial\sigma(\varphi
,T)/\partial\varphi)_{T}[(\partial\varphi/\partial T)+S(\varphi,T)]\\
&  +\sigma(\varphi,T)(\partial S(\varphi,T)/\partial\varphi)_{T}.
\end{align*}
We now differentiate the first condition in (\ref{Equilibriumphi}) and use the
above equation at equilibrium to obtain%
\begin{align*}
&  (\partial\sigma(\overline{\varphi}_{\text{b}},T)/\partial\overline{\varphi
}_{\text{b}})_{T}[(\partial\overline{\varphi}_{\text{b}}/\partial
T)+S(\overline{\varphi}_{\text{b}},T)]\\
&  =-\beta^{2}-\beta(\partial S(\overline{\varphi}_{\text{b}},T)/\partial
\overline{\varphi}_{\text{b}})_{T}.
\end{align*}
The basin entropy $S(\varphi,T)$ can also be expressed as $S(\mathcal{S}%
,T),$\ so that the basin free energy function can be written as $\varphi
(\mathcal{S},T)=E(T)-TS(\mathcal{S},T).$ From this, we obtain%
\[
(\partial\varphi(\mathcal{S},T)/\partial\mathcal{S)}_{T}=-T(\partial
S(\mathcal{S},T)/\partial\mathcal{S)}_{T}.
\]
At equilibrium ($\mathcal{S=}\overline{\mathcal{S}}$), the left-hand side is
equal to $T$ from the first condition in (\ref{Equilibriumphi}). Thus, at
equilibrium,
\[
(\partial S(\overline{\mathcal{S}},T)/\partial\overline{\mathcal{S}%
}\mathcal{)}_{T}=-1.\
\]
Since $(\partial S(\overline{\varphi}_{\text{b}},T)/\partial\overline{\varphi
}_{\text{b}})_{T}=(\partial S(\overline{\mathcal{S}},T)/\partial
\overline{\mathcal{S}}\mathcal{)}_{T}(\partial\overline{\mathcal{S}}%
/\partial\overline{\varphi}_{\text{b}}\mathcal{)}=-(\partial\mathcal{S}%
(\overline{\varphi}_{\text{b}},T)/\partial\overline{\varphi}_{\text{b}%
}\mathcal{)}=-[\beta+(\partial\mathcal{S}/\partial T\mathcal{)}_{\overline
{\varphi}_{\text{b}}}(\partial T/\partial\overline{\varphi}_{\text{b}%
})]=-\beta\lbrack1+S_{\text{b}}(T)(\partial T/\partial\overline{\varphi
}_{\text{b}})],$ where we have we used (\ref{ComplexityCoefiiciet1}) at
equilibrium [$\sigma(\overline{\varphi}_{\text{b}},T)=\beta,$ and
$S_{\text{b}}(T)=S(\overline{\varphi}_{\text{b}},T)]$. Thus, we find that
\begin{align*}
&  (\partial\sigma(\overline{\varphi}_{\text{b}},T)/\partial\overline{\varphi
}_{\text{b}})_{T}[(\partial\overline{\varphi}_{\text{b}}/\partial
T)+S_{\text{b}}(T)]\\
&  =\beta^{2}S_{\text{b}}(T)(\partial T/\partial\overline{\varphi}_{\text{b}%
}),
\end{align*}
so that the second condition of stability becomes%
\begin{align*}
(\partial\sigma(\overline{\varphi}_{\text{b}},T)/\partial\overline{\varphi
}_{\text{b}})_{T}  &  =\beta^{3}S_{\text{b}}(T)(\partial T/\partial
\overline{\varphi}_{\text{b}})/\left(  \partial\overline{\mathcal{S}}/\partial
T\right) \\
&  <0,
\end{align*}
where we have used the relation
\begin{equation}
T\left(  \partial\overline{\mathcal{S}}/\partial T\right)  =\partial
\overline{\varphi}_{\text{b}}/\partial T+S_{\text{b}}(T)
\label{ComplexityBasinRelation}%
\end{equation}
obtained from (\ref{DifferentialPhi}). Therefore, the second condition of
stability finally becomes%
\[
\left(  \partial\overline{\mathcal{S}}/\partial T\right)  (\partial
\overline{\varphi}_{\text{b}}/\partial T)<0.
\]
We expect $\left(  \partial\overline{\mathcal{S}}/\partial T\right)  $ to be
positive at low temperatures, so the stability condition there reduces to
$(\partial\overline{\varphi}_{\text{b}}/\partial T)<0.$ Because of
(\ref{ComplexityBasinRelation}), it is easy to see that $\left(
\partial\overline{\mathcal{S}}/\partial T\right)  >0$ even if $(\partial
\overline{\varphi}_{\text{b}}/\partial T)$ changes sign, so the stability
always requires
\[
\left(  \partial\overline{\mathcal{S}}/\partial T\right)  >0;~(\partial
\overline{\varphi}_{\text{b}}/\partial T)<0.
\]
\ 

It is our pleasure to thank Fedor Semerianov for various discussions.

\end{document}